\documentclass[10pt]{iopart}
\usepackage{iopams}
\usepackage[utf8]{inputenc}
\usepackage{color}
\usepackage{overpic}
\usepackage{soul}
\usepackage{cite}

\newcommand{\DREAM}{\textsc{Dream}}
\usepackage{graphicx}
\usepackage[numbers]{natbib}
\newcommand{\eqref}[1]{(\ref{#1})}

\usepackage{float}
\usepackage{graphicx}
\graphicspath{{./figures/}}
\usepackage[export]{adjustbox}

\usepackage{pgfplots}
\pgfplotsset{compat=1.15}
\usepackage{pgfplotstable} 
\usepgfplotslibrary{external} 
\usepackage[breaklinks=true]{hyperref}
\hypersetup{
  unicode=false,            pdftoolbar=true,          pdfmenubar=true,          pdffitwindow=false,       pdfstartview={FitH},      pdfsubject={Subject},     pdfcreator={Creator},     pdfproducer={Producer},   pdfkeywords={keyword1} {key2} {key3},   pdfnewwindow=true,        colorlinks=true,          linkcolor=blue,           citecolor=blue,           filecolor=blue,           urlcolor=blue           }

\begin{document}

\title[O.~Vallhagen et al, RE Dynamics in ITER Disruptions with SPI]{Runaway Electron Dynamics in ITER Disruptions with Shattered Pellet Injections
  }
\author{O.~Vallhagen$^1$, L.~Hanebring$^1$, F.~J.~Artola$^{2}$, M.~Lehnen$^2$, E.~Nardon$^3$, T.~F\"ul\"op$^1$, M.~Hoppe$^4$, S.~L.~Newton$^5$, I.~Pusztai$^1$}
\address{$^1$ Department of Physics, Chalmers University of Technology,
  SE-41296 Gothenburg, Sweden}
\address{$^2$ITER Organization, Route de Vinon-sur-Verdon, CS 90 046, 13067 St. Paul Lez Durance Cedex, France}
\address{$^3$CEA, IRFM, F-13108 Saint-Paul-lez-Durance, France}
\address{$^4$Department of Electrical Engineering, KTH Royal Institute of Technology, SE-11428 Stockholm, Sweden}
\address{$^5$United Kingdom Atomic Energy Authority, Culham Science Centre, Abingdon, Oxon OX14 3DB, United Kingdom}
    \ead{vaoskar@chalmers.se}
\begin{abstract}
This study systematically explores the parameter space of disruption mitigation through shattered pellet injection in ITER with a focus on runaway electron dynamics, using the disruption modelling tool \DREAM. The physics fidelity is considerably increased compared to previous studies, by e.g., using realistic magnetic geometry, resistive wall configuration, thermal quench onset criteria, as well as including additional effects, such as ion transport and enhanced runaway electron transport during the thermal quench. The work aims to provide a fairly comprehensive coverage of experimentally feasible scenarios, considering plasmas representative of both non-activated and high-performance DT operation, different thermal quench onset criteria and transport levels, a wide range of hydrogen and neon quantities injected in one or two stages, and pellets with various characteristic shard sizes. Using a staggered injection scheme, with a pure hydrogen injection preceding a mixed hydrogen-neon injection, we find injection parameters leading to acceptable runaway electron currents in all investigated discharges without activated runaway sources. Dividing the injection into two stages is found to significantly enhance the assimilation and minimize runaway electron generation due to the hot-tail mechanism. However, while a staggered injection outperforms a single stage injection also in cases with radioactive runaway electron sources, no cases with acceptable runaway electron currents are found for a DT-plasma with a $15\,\rm MA$ plasma current.     
\end{abstract}

\ioptwocol

\section{Introduction}
Disruptions -- off-normal events when the plasma energy is rapidly lost -- and associated generation of highly energetic runaway electron (RE) beams represent an outstanding challenge of the tokamak concept for magnetic confinement. An effective disruption mitigation system in a tokamak reactor should limit the exposure of the wall to localised heat losses during the thermal quench (TQ), and to the impact of RE beams, as well as avoid excessive forces on structural elements. In an ITER plasma at $15\,\rm MA$ plasma current, avoiding excessive electromagnetic forces requires the current quench (CQ) time to be kept in the range $50$--$150\,\rm ms$, and the RE current should be kept below $150\,\rm kA$ \cite{LehnenITER}. 

The currently envisaged mitigation method is to inject a massive amount of material, primarily consisting of a mixture of neon (Ne) and hydrogen (H), when an emerging disruption is detected, in order to better control the plasma cooling and energy dissipation. In ITER, the baseline strategy for delivering the material is through shattered pellet injection (SPI) \cite{LehnenITER}. However, finding operation parameters in reactor-scale devices that simultaneously avoid RE formation and mitigate heat and electromagnetic loads remains an open question. 

Several previous studies of integrated simulations of the full disruption event have been made with a focus on the CQ evolution, based on simplified geometry and/or assumptions of a prescribed material deposition and temperature drop \cite{Martin2017Formation,Vallhagen2020Runaway,Pusztai2023Bayesian}. Although earlier results \cite{Martin2017Formation} indicated that the RE generation could be suppressed by a sufficiently large H injection, later results suggested that the recombination occurring at very high H quantities counteracts the damping of the runaway generation, thereby boosting the avalanche RE generation \cite{Vallhagen2020Runaway,Pusztai2023Bayesian}, leading to several MAs of RE currents in all studied cases of ITER-like DT plasmas. 

An initial study of simulations of SPI-mitigated disruptions in  ITER was presented in \cite{Vallhagen2022Effect}, using a self-consistent model for the injection and temperature drop for a given evolution of the magnetic perturbation amplitude. The main objective of this paper was to investigate the effect of separating the injection into two phases, as previously considered in \cite{Nardon2020Fast}. The resulting two-stage cooling was found to effectively thermalize the tail of the electron distribution function before the onset of the RE generation, strongly reducing the RE seed due to the hot-tail mechanism. This indicated a possibility to obtain low RE currents in plasmas without tritium decay and Compton scattering of  photons from the radioactive wall (referred to as activated sources). Nonetheless, large RE currents were still found for DT plasmas. However, these simulations studied a rather limited set of scenarios and still included several simplifications, such as circular flux surface cross sections, a fixed TQ onset time (independent of the evolution of the plasma parameters), and neglected ion transport once the injected material was ablated. 

In this study we expand the parameter space as well as the range and complexity of the scenarios compared to the previous studies, while also improving the physics fidelity. The simulations are performed with the disruption simulation tool \DREAM{} \cite{Hoppe2021DREAM}. The \DREAM{} code is designed to include much of the relevant physics during a disruption while keeping the complexity sufficiently low (e.g., by resolving one spatial dimension) to make it feasible to explore a large parameter space, or even allowing for optimization \citep{Pusztai2023Bayesian}. It includes an SPI ion source based on an analytical ablation rate \cite{Vallhagen2022Effect}, time dependent rate equations for ionisation and radiation, as well as a self-consistent evolution of the induced electric field. 

\DREAM{} covers a wide range of the computational fidelity-affordability trade-off concerning RE physics, up to a linearized bounce-averaged relativistic Fokker-Planck treatment. Here, we use a computationally inexpensive fluid RE model, allowing the exploration of a large parameter space. Further details about the model and the simulation settings are covered in section \ref{sec:settings} and the simulation results are described in section \ref{sec:results}. The results and limitations of the model are discussed in section \ref{sec:discussion}, before the paper is concluded in section \ref{sec:conclusion}.

\section{Simulation Settings}
\label{sec:settings}
\begin{figure*}
    \centering
    \includegraphics[width=0.49\textwidth]{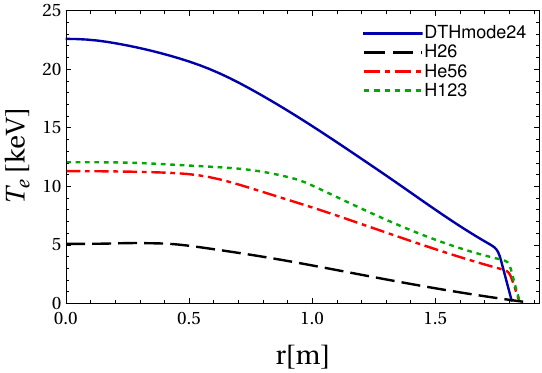}
    \put(-200,150){a)}
    \includegraphics[width=0.49\textwidth]{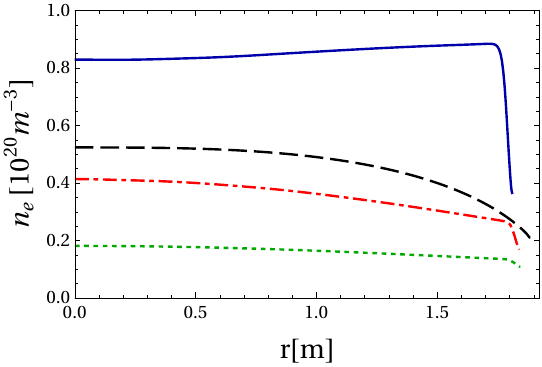}
    \put(-200,145){b)}
    \begin{flushleft}
    \includegraphics[width=0.5\textwidth]{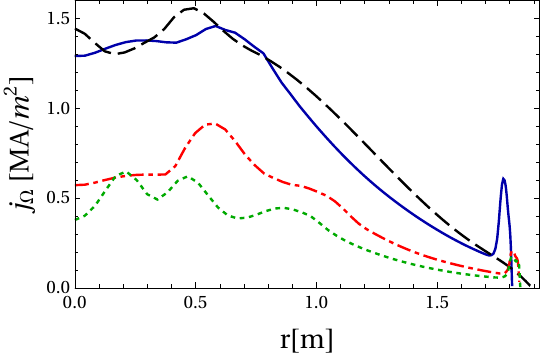}
    \put(-200,150){c)}
    \hspace{3cm}
    \includegraphics[width=0.25\textwidth]{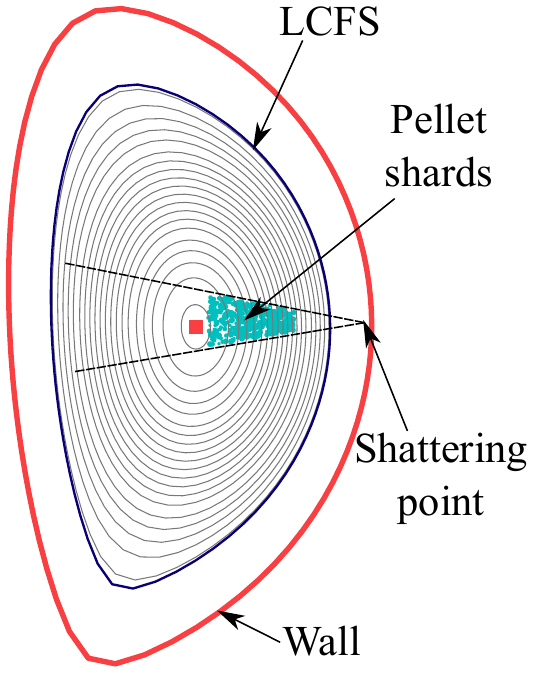}
    \put(-130,150){d)}
    \end{flushleft}
    \caption{a-c) Initial plasma parameter profiles for the DTHmode24 (solid), H26 (dashed), He56 (dash-dotted) and H123 (dotted) scenarios. a) Electron temperature. b) Electron density. c) Current density. d) Flux surface and SPI geometry for the H26 scenario (The other scenarios are similar).}
    \label{fig:profiles}
\end{figure*}

The simulations are based on four reference scenarios produced by the CORSICA \cite{Kim_2018} workflow, which are commonly used within the ITER Organisation. These scenarios are referred to as {\em DTHmode24}, a high-performance H-mode D-T plasma (pure, and of equal isotope concentrations) with a $15\,\rm MA$ plasma current; {\em H26}, a hydrogen L-mode plasma with a $15\,\rm MA$ plasma current; {\em He56}, a helium H-mode plasma with a $7.5\,\rm MA$ plasma current; and {\em H123}, a hydrogen H-mode plasma with a $5\,\rm MA$ plasma current. The flux surface geometry and initial profiles for the studied cases are illustrated in Fig.~\ref{fig:profiles}. Moreover, we assume a resistive wall time of $t_\mathrm{wall}=0.5\,\rm s$ and a wall radius of $r_\mathrm{wall}=2.833\,\rm m$, chosen to match the poloidal field energy within the vacuum vessel (which is the toroidally closed conductor closest to the plasma) in JOREK \cite{Huysmans2007MHD} simulations, corrected for contributions from the poloidal field coils.

\subsection{SPI parameters}
\label{sec:SPI}
The hydrogen isotope of pellets used in our simulations is deuterium (D) which is expected to produce very similar results to H, the isotope planned to be employed in ITER. The reason for this choice is that the pellet ablation model we use is developed for D. The pellet injection speed is assumed to be $v_\mathrm{p}=500\,\rm m/s$, and the fragment velocity dispersion is uniform within $v_\mathrm{p}\pm \Delta v$, with $\Delta v/v_\mathrm{p}=0.4$. The injection is modelled in a poloidal plane within a spreading angle of $\alpha=10^\circ$, with the shards diverging from the point $\{R,Z\}=\{8.568,0.6855\}\,\rm m$ (with the origin at the horizontal midplane at the center of the torus), as shown in Fig.~\ref{fig:profiles}d. We take into account a $B$-field correction of the ablation, representing the diamagnetic shielding of the pellet \cite{Samulyak2021Lagrangian}.

The SPI fragment sizes follow the distribution proposed by Parks \cite{Parks2016Modeling}, and we consider three different characteristic shard sizes, corresponding to 487 (default shard size), 5185 (small shard size) and 68 (large shard size) shards for a standard ITER pellet (consisting of $1.85\cdot 10^{24}$ atoms), as also done in previous INDEX simulations \cite{AkinobuITPA}. As the Ne concentration is varied, the D quantities are adjusted to keep the total number of atoms in the pellet constant. We consider both single and multiple pellet injections with up to four pellets, injected in either one or two stages.    

We consider two Ne concentrations for each scenario, chosen to roughly span the desired ohmic CQ time range of $\sim 50-150 \,\rm ms$, to the extent possible (see related discussion about Fig.~\ref{fig:correlations_current}a in section \ref{sec:correlations_current}). The ohmic CQ time is defined here as $t_\mathrm{CQ}=\left[t(I_\mathrm{Ohm}=0.2I_\mathrm{p0})-t(I_\mathrm{Ohm}=0.8I_\mathrm{p0})\right]/0.6$, where $I_\mathrm{Ohm}$ is the total ohmic current, and $I_\mathrm{p0}$ is the initial total plasma current. For each case $t_\mathrm{CQ}$ is evaluated based on an additional simulation with runaway generation disabled but otherwise identical settings. The choice of Ne concentrations is based on initial tests on a baseline case, using a single stage injection, the default shard size and a TQ of $3\,\rm ms$ duration with the ``late'' onset criterion, (see section \ref{sec:TQ}). The chosen Ne quantities are $N_\mathrm{Ne,inj}=2\cdot 10^{23}$ and $N_\mathrm{Ne,inj}=1.83\cdot 10^{24}$ for the H26 scenario; $N_\mathrm{Ne,inj}=2.5\cdot10^{22}$ and $N_\mathrm{Ne,inj}=1.5\cdot 10^{24}$ for the DTHmode24 scenario; $N_\mathrm{Ne,inj}=5\cdot 10^{21}$ and $N_\mathrm{Ne,inj}=1.83\cdot 10^{24}$ for the He56 scenario; and $N_\mathrm{Ne,inj}= 10^{21}$ and $N_\mathrm{Ne,inj}=1.83\cdot 10^{24}$ for the H123 scenario. 
Additionally, for every plasma scenario considered, we perform simulations using an injected Ne content of $N_\mathrm{Ne,inj}=5\cdot10^{22}$, to assist comparison between scenarios, as well as to connect the results to previous work with INDEX \cite{AkinobuITPA}.

The cloud of pure hydrogenic pellets is prone to build an excess pressure and thus drift significant distances towards the low field side \cite{Parks2000Radial,Rozhansky2004Mass,Pegourie2006Homogenization}. Although a computationally efficient model to self-consistently capture this effect has been developed \cite{Vallhagen2023Drift}, it is not yet available in \DREAM. Instead, to emulate this effect in two stage injections modelled here, the material deposition of the first set of pellets (with no neon content) is shifted one radial grid cell (with a width of about $20\,\rm cm$) outward of the shard position\footnote{To avoid additional free parameters and model complexity, we neglect the heat transport resulting from the displacement of the material. We did however perform sample simulations where all the heat passing through a radius of $1\rm\, cm$ around the shard was displaced by the same distance as the material, and found that this heat transport only had a minor impact on the result.}. Then the pellet shard plume experiences a considerably reduced dilution cooling generated by the material deposition, leading to an outward-shifted, sometimes quite edge-localised, deposition profile. To bracket the range of drift effects, for comparison we also perform similar simulations with local deposition.

The ablated material is initially deposited into the neutral charge state, and the ionization is then evolved using time dependent rate equations. The ionization and recombination rates, as well as the corresponding radiation rates, are taken from the ADAS database \cite{ADAS} for Ne and the AMJUEL database\footnote{http://www.eirene.de/html/amjuel.html} for D. The latter accounts for opacity to Lyman radiation, which has previously been found to be important at the high injected D densities studied here \cite{Vallhagen2022Effect}.

\subsection{Thermal quench}
\label{sec:TQ}
We consider two thermal quench onset criteria. According to the \emph{early} onset criterion, the TQ is triggered when a Ne containing shard with a horizontal velocity of $v_p$ reaches the $q=2$ surface. In the \emph{late} onset criterion the TQ is triggered when any point inside of the $q=2$ flux surface reaches a temperature of $10\,\rm eV$; in case of pellets with low neon concentration this criterion is usually, although not always, satisfied significantly later than the early one – hence the name. 

These criteria are motivated by the point that the resistive decay of the current profile is sufficiently fast at $10\,\rm eV$ to likely trigger a magnetohydrodynamic (MHD) instability around the $q=2$ flux surface on a millisecond time scale. The temperature drop to $10\,\rm eV$, however, might occur much earlier locally (i.e., somewhere within the flux surface) than on flux surface average. In particular, such a local cooling might happen in the vicinity of  Ne containing shards, which, in an extreme case, might be sufficient to trigger an MHD instability as the shards pass the $q=2$ flux surface. This scenario motivates the early onset criterion, while the late onset criterion corresponds to the opposite extreme case, where the temperature variation within the flux surface remains negligible. These two criteria are thus intended to bracket the range of possible TQ onset delays.

Once the TQ is triggered, spatially and temporally homogeneous electron heat diffusion and runaway electron particle diffusion are activated, and remain active for a period of $t_\mathrm{TQ}=1\,\rm ms$ or $3\,\rm ms$ in different scenarios. These time scales are in the range indicated by extrapolations of empirical data from present day devices, and can be considered as typical values used to assess thermal loads in an ITER TQ \cite{loarteTQ}.
The diffusion coefficients are parameterized by a relative magnetic perturbation amplitude, $\delta B/B$, using a Rechester-Rosenbluth-type model \cite{Hoppe2021DREAM,Rechester1978Electron} for the heat transport, and the model by Svensson et al.~\cite{Svensson2021Effects} for RE transport.  The values of $\delta B/B$ are chosen such that the temperature at the magnetic axis would have time to drop below $200\,\rm eV$ in the time $t_\mathrm{TQ}$ if transport would be the only loss mechanism (note though, that it typically reaches much lower values due to radiative energy losses). At lower temperatures, Ohmic heating can balance the heat losses due to transport, preventing the temperature from dropping much further due to transport alone. The resulting values of $\delta B/B$ are tabulated in \ref{app:dBB}, and are of the order of $0.1$--$1\%$ –- comparable to those extracted from an MHD simulation of an SPI triggered ITER TQ, in \cite{Hu2021Radiation}. 

Ion particle transport (of all ion species and charge states) is activated at the same time as the other transport channels, with initial diffusion and advection coefficients of $D_\mathrm{ion}=4000\,\rm m^2/s$ and $A_\mathrm{ion}=-2000\,\rm m/s$, which then decay exponentially over the characteristic time of $\tau=0.5\,\rm ms$. A similar form of the time evolution of the ion transport coefficients was used in \cite{Linder2020Self} to reproduce experimental data in the ASDEX Upgrade tokamak with the ASTRA code. The chosen values of $D_\mathrm{ion}$ and $A_\mathrm{ion}$ give diffusion and advection time scales ($r_\mathrm{minor}^2/D_\mathrm{ion}$ and  $r_\mathrm{minor}/A_\mathrm{ion}$, respectively, where $r_\mathrm{minor}$ is the minor radius of the plasma) in the ms range. Together they result in a transport of a substantial amount of material to the plasma core over $\sim 0.1\,\rm ms$, as expected from 3D MHD simulations \cite{Hu2021Radiation}, and the balance between the advection and the diffusion gives a moderately peaked final profile ($r_\mathrm{minor}/L_n\approx -r_\mathrm{minor} A_\mathrm{ion}/D_\mathrm{ion}\approx 1$ where $L_n$ is the characteristic density length scale). The decay time constant $\tau=0.5\,\rm ms$ ensures that the ion transport is significant on the same time scale as the other transport processes are active.

\subsection{Runaway electrons}
\label{sec:RE}
The \DREAM{} simulations use the fully fluid runaway model. In simulations of the H26, He56 and H123 scenarios, only non-activated runaway seed generation mechanisms are included: the Dreicer generation uses a neural network trained on kinetic simulations\footnote{The Dreicer runaway generation rate is calculated using an analytical extrapolation at very small electric fields, outside the training range of the neural network, to avoid an overestimation of the corresponding seed.} \cite{NN_Dreicer}, and the hot-tail model used is described in Appendix C in \cite{Hoppe2021DREAM}. The avalanche multiplication rate is described in \citep{Hesslow_fluid_ava}, and it accounts for both partial screening and magnetic trapping effects.   The Svensson model \cite{Svensson2021Effects} is used to turn the momentum $p$ and pitch $\xi=p_\|/p$ dependent diffusion coefficients into diffusion coefficients of the runaway fluid. Here the assumed $p$-dependence is $p/(1+p^2)$, that reproduces the $\propto v$ low energy behavior, and emulates a reduction of the runaway transport due to finite Larmor radius and orbit width effects at high $p$. The  $\xi$-dependence corresponds to the $\propto|v_\||$-dependence of the Rechester-Rosenbluth type diffusion coefficients. In addition, the DTHmode24 case also includes RE generation by activated sources. 

According to \cite{Martin2017Formation}, the photon flux from the wall is expected to drop by a factor of $\sim 1/1000$ immediately after the neutron bombardment of the wall is stopped (i.e., when the fusion reaction ceases). To account for this in the calculation of the Compton seed, we start the simulations with the nominal photon flux on ITER given in \cite{Martin2017Formation}, and then reduce it at the end of the TQ (taken to be the end of the transport event) by a factor of $1/1000$. The ITER nominal photon flux taken from \cite{Martin2017Formation} was calculated assuming a beryllium wall design.

\section{Simulation results}
\label{sec:results}

The difficulty of avoiding a large RE formation is strongly dependent on the initial plasma current, due to the maximum RE avalanche gain being exponentially sensitive to it. We therefore study the scenarios with a $15\,\rm MA$ plasma current separately in section \ref{sec:example} and \ref{sec:correlations}, before turning to the cases with reduced plasma current in section \ref{sec:reduced_Ip}.

\subsection{Representative example}
\label{sec:example}

\begin{figure*}[h!]
    \centering
    \includegraphics[trim={0 1.62cm 0 0},clip,width=0.32\textwidth]{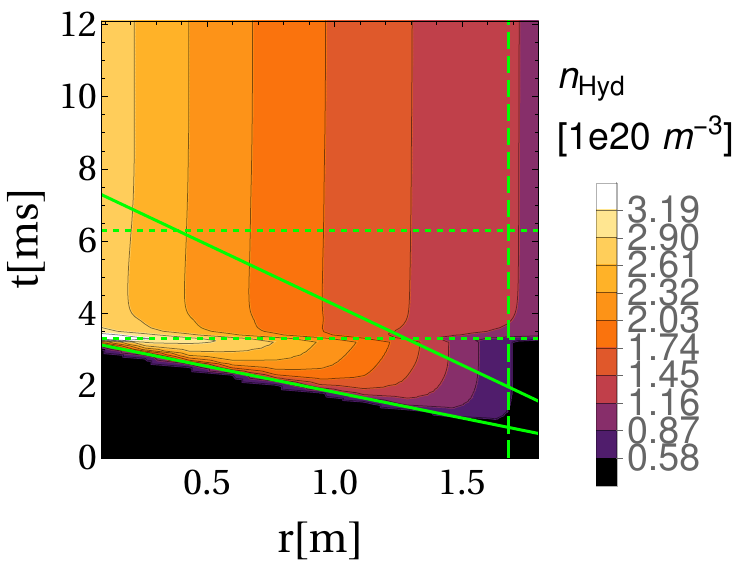}
    \put(-126,80){\textcolor{black}{a)}}
    \includegraphics[trim={0 1.62cm 0 0},clip,width=0.30\textwidth]{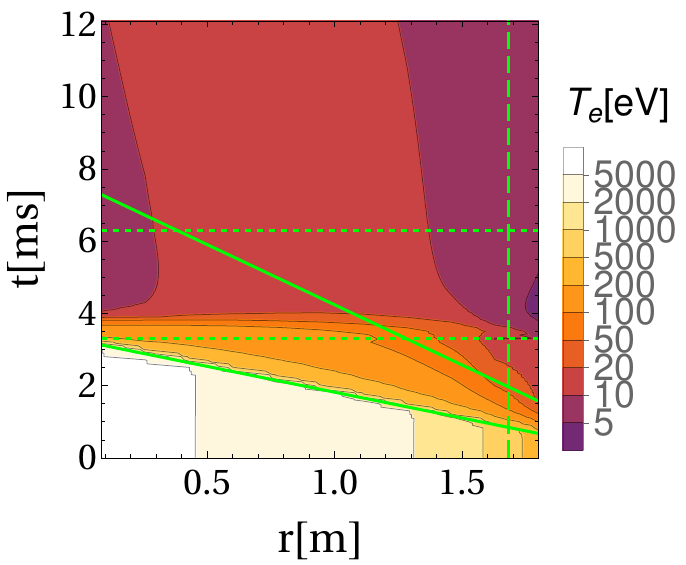}
    \put(-115,80){\textcolor{white}{b)}}
    \includegraphics[trim={0 1.0cm 0 0},clip,width=0.34\textwidth]{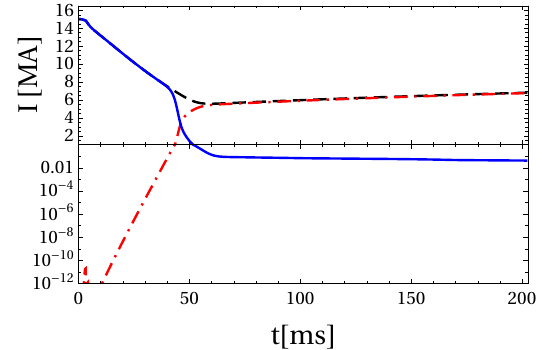}
    \put(-126,77){\textcolor{black}{c)}}
    
    \includegraphics[width=0.32\textwidth]{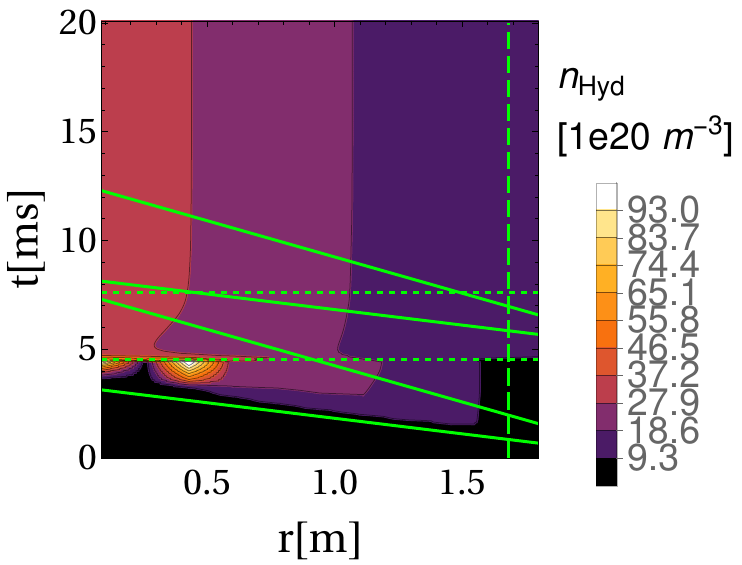}
    \put(-130,100){\textcolor{white}{d)}}
    \includegraphics[width=0.30\textwidth]{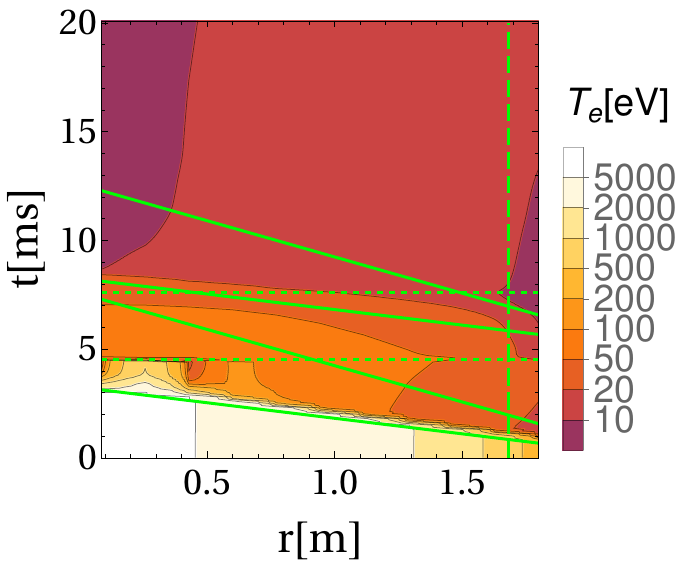}
    \put(-120,103){\textcolor{white}{e)}}
    \includegraphics[width=0.34\textwidth]{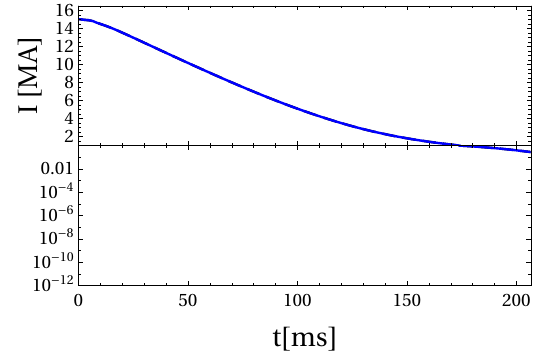}
    \put(-115,95){\textcolor{black}{f)}}
    \caption{Plasma parameter evolutions for a baseline H26 case with a target CQ time of 100 ms. a-c) single stage injection, and d-f) staggered injection with a shifted deposition for the first injection. a) and d) Hydrogen density (injected+background). b) and e) Electron temperature. c) and f) Time evolution of the total plasma current (dashed), and its ohmic (solid) and runaway (dash-dotted) components. In a-b) and d-e) solid green lines indicate the trajectory of the fastest and slowest shards for each injection, the dashed line marks the $q=2$ flux surface, and the temporal boundaries of the transport event are shown with dotted lines.}
    \label{fig:s2_st2}
\end{figure*}

We start by considering one of the baseline cases (late TQ onset criterion and a $3\,\rm ms$ TQ) with the H26 scenario, which serves as a representative example of a single stage injection. Fig.~\ref{fig:s2_st2} shows the time evolution of the most important quantities for the case when the neon quantity is adjusted ($10.8\%$) to achieve a CQ time of $100\,\rm ms$. The total hydrogenic ion density increases along the trajectory of the fastest shards (lower solid green line), as shown in Fig.~\ref{fig:s2_st2}a. After the ions are rapidly mixed across the entire radius in the beginning of the transport event (indicated by the lower dotted line),  the density remains steady for the rest of the simulation. Note that in single pellet injection cases the density evolution of neon is very similar, within the multiplicative factor of the injected Ne/D ratio. We also note that the assimilated fraction of the injected material in this case is $f_\mathrm{assim}=5.18\%$, which corresponds to $1.04\cdot 10^{22}$ assimilated neon atoms and $8.55\cdot 10^{22}$ hydrogen atoms.

In Fig.~\ref{fig:s2_st2}b, we can see the temperature drop below the keV-range after the first shards reach the plasma, and once the temperature locally goes below $10\,\rm eV$ just inside the $q=2$ flux surface, the transport event is initiated. This leads to a more homogeneous further cooling of the plasma, ending at $T_e\sim 10\,\rm eV$. The initially created RE seed that follows the first cooling gets mostly lost in the transport event, then the RE current recovers and increases slowly and reaches macroscopic values around $t=40\,\rm ms$, as shown with the dash-dotted line of Fig.~\ref{fig:s2_st2}c. Note that the RE current at the end of the transport event is tiny, $\sim 10^{-10}\,\rm A$ (outside the scale of figure \ref{fig:s2_st2}c), and it mostly stems from a hot-tail seed. A small, but non-negligible, Dreicer seed generation remains active during the earlier parts of the CQ, generating a representative seed of $\sim 10^{-8}\,\rm A$, but the seed still remains very small. 

The representative RE current is still significant, $5.31\,\rm MA$, due to the very strong avalanche. We introduced two quantities; the \emph{representative seed} is the total RE current at the end of the TQ (that is not lost due to transport) plus the seed generated after the TQ; and the \emph{representative RE current} is RE current when $95\%$ of the total current is carried by REs. Henceforth these are understood when we refer to a scalar RE current or seed without further qualifications.

An important general feature -- which is more pronounced in the well performing cases -- is that the temperature drop is usually divided into two separate stages, despite all material being injected in a single pellet. This can effectively suppress the hot-tail seed generation, as this enables the hot-tail to slow down and thermalize with the bulk at an intermediate temperature before a strong electric field is induced after the second, radiatively dominated, cooling phase. 
In addition, in a late TQ the significantly higher density increase, which is allowed by a higher temperature, also assists thermalization. 

The separation of the cooling stages has also been observed in simulations with the INDEX and JOREK codes \cite{Matsuyama_2022}, and is explained by the temperature dependence of the various heat loss mechanisms. Transport as a heat loss channel becomes inefficient in the $100\,\rm eV$ range and below, while radiative cooling due to the presence of neon has a local minimum around $200\,\rm eV$ and then rises sharply towards lower temperatures (see e.g.~Fig.~6 in \cite{Vallhagen2022Effect}). During the injection phase, the temperature is also strongly affected by dilution. Due to the 1D nature of the employed model, the flux surface averaged dilution cooling occurs instantaneously after deposition, thus the effect of dilution on the temperature is determined directly by the ablation rate. The strong scaling of the ablation rate with temperature makes this process dominant in the early stage of the injection (before the onset of the transport event), while this process also becomes inefficient in the $100\,\rm eV$ range and below.

In case of a late thermal quench, the cooling starts with a dilution down to a few hundred $\rm eV$, and then it slows down due to the low radiated power. Eventually, the temperature drops below $100\,\rm eV$ in some parts of the plasma where cooling accelerates again due to radiation. As the temperature drops below $10\,\rm eV$ at some radial location,  the transport event is triggered. Transport-induced cooling, although remaining moderate in this temperature range, helps drive the rest of the plasma below $100\,\rm eV$, initiating a global radiative collapse. This global radiative collapse is further assisted by the particle transport, increasing dilution, as well as the radiating neon content, in regions with low or no direct deposition of injected material.

The separation of the cooling phases is typically less pronounced in cases with an early TQ, and also in cases with large neon injections, typically resulting in larger RE currents. In those cases, the earlier cooling also reduces the ablation, and hence the frictional drag on energetic electrons, which further enhances the RE generation. Indeed, the scenario shown in Fig.~\ref{fig:s2_st2}a-c when simulated with an early TQ gave a representative RE current of $7.72\,\rm MA$. When aiming for a CQ time of $50\,\rm ms$ (which in this case means a pellet consisting of 99\% neon, see section \ref{sec:SPI}), we obtained a representative RE current of $11.7\,\rm MA$. 

On the other hand, both the separation of the cooling phases and the collisional damping of RE generation are more pronounced with a staggered injection. In those cases, the lack of radiating neon during the first injection stage allows for a higher temperature in the vicinity of the shards, resulting in a higher ablation, and also delays the onset of the transport event and the radiative collapse with both the early and late TQ onset criterion. An example of such an evolution is shown in Fig.~\ref{fig:s2_st2}d-f for a similar case to that in Fig.~\ref{fig:s2_st2}a-c but with the neon-containing pellet preceded by a full deuterium pellet by $5\,\rm ms$. Notably, in this case the RE seed becomes negligible, and hence no RE current is generated.

The drift of the ablated material can lead to a shift in the deposition profile from the core towards the edge. This shift can be much larger than the drift length itself, since as the dilution cooling caused by the shards is displaced behind the plume of shards, the higher temperature in the vicinity of the shards greatly enhances their ablation. Thus the shards may ablate away rapidly, before reaching the core.

In the example shown in Fig.~\ref{fig:s2_st2}d, where the shift between the ablation location and the deposition location is only $\sim 20\,\rm cm$, the outward shift of the deposition profile is moderate, which is due to the moderate temperature of the background plasma of the H26 scenario. In contrast, in the $\sim 4$ times hotter DTHmode24 plasma we sometimes observe deposition profiles peaked as far out as $r\sim 1.5 \,\rm m$, for the same deposition shift of $\sim 20\,\rm cm$ (not shown here). The deposition may in some cases be so localized, that the corresponding dilution cooling may be sufficient to meet our \emph{late} TQ trigger condition, even before the neon doped shards arrive into the plasma. This is the case in the example shown in Fig.~\ref{fig:s2_st2}e, where the transport onset (lower dotted line) appears at an earlier time than the second injection (upper pair of solid lines).    

In spite of these, potentially dramatic, differences in the early evolution of the disruption between local and shifted material deposition, the effects of the shifted deposition are reduced by the transport event. Namely, the density profile is strongly flattened by the fast ion mixing during the transport event. As a result, the CQ evolutions with and without the shift included are rather similar. We should keep in mind though, that our results rely on an effective ion mixing; the efficiency of this process is a source of uncertainty. Moreover, if the pellet cloud drift would be comparable to the plasma minor radius it could cause a significant amount of material to be completely ejected from the plasma, which might have a significant effect on the CQ evolution. It is worth noting that within our model, assuming that no such material ejection takes place, the shift also leads to a somewhat increased amount of assimilated D in the plasma, as will be discussed in connection to Fig.~\ref{fig:correlations_assim}. 

\subsection{Correlation among the explored parameters with a $15\,\rm MA$ plasma current}
\label{sec:correlations}

In this section we study correlations between the various settings and figures of merit covered in this study, by plotting the various settings and metrics against each other, displaying pairs of quantities with observable trends of interest. A list of the simulation settings included is provided in \ref{app:settings}. Section \ref{sec:correlations_current} focuses on quantities related to the plasma current evolution, while section \ref{sec:correlations_assim} focuses on quantities related to the assimilation of the injected material.

\subsubsection{Current evolution}
\label{sec:correlations_current}

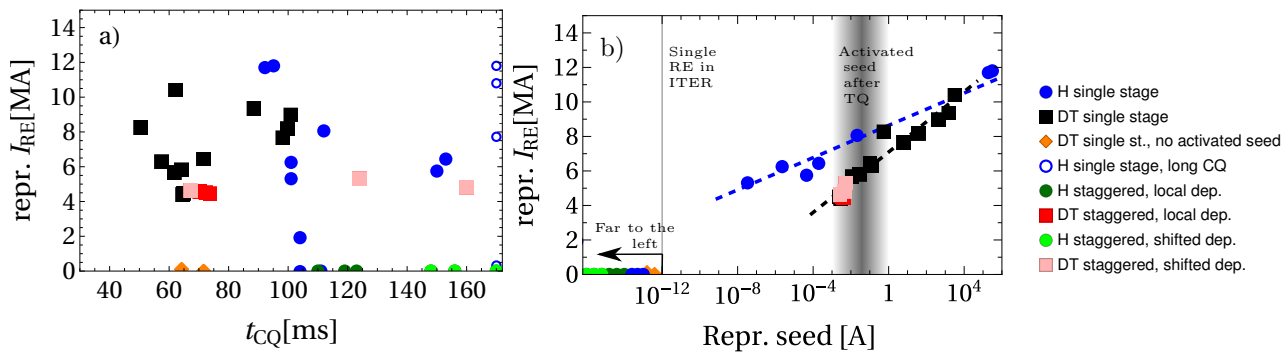
\begin{figure*}
    \begin{minipage}[h]{0.79\textwidth}
    \includegraphics[width=0.49\textwidth]{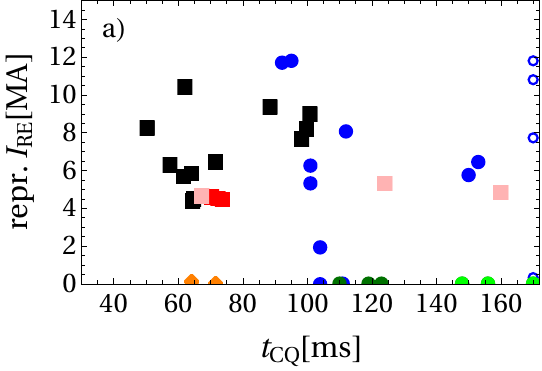}
    \includegraphics[width=0.48\textwidth]{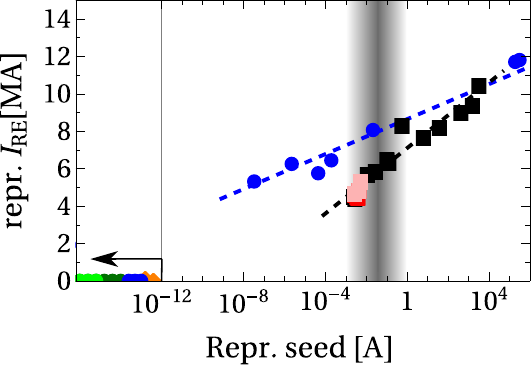}
        \put(-62,112){\tiny \textcolor{black}{Activated}}
        \put(-60,106){\tiny \textcolor{black}{seed}}
        \put(-60,100){\tiny \textcolor{black}{after}}
        \put(-60,94){\tiny \textcolor{black}{TQ}}
        \put(-126,112){\tiny \textcolor{black}{Single}}
        \put(-126,106){\tiny \textcolor{black}{RE in}}
        \put(-126,100){\tiny \textcolor{black}{ITER}}
        \put(-154,45){\tiny \textcolor{black}{Far to the}}
        \put(-140,39){\tiny \textcolor{black}{left}}
        \put(-152,112){\textcolor{black}{b)}}
    \end{minipage}
    \begin{minipage}[h]{0.2\textwidth}
    \includegraphics[width=\textwidth]{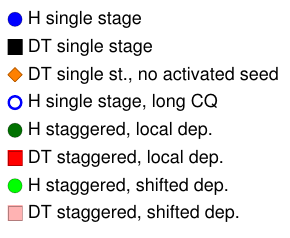}
        \end{minipage}
    \caption{Correlations between figures of merit related to the plasma current evolution, varying all the simulation parameters within the ranges described in section~\ref{sec:settings}. a) Representative RE current vs. CQ time. Cases with incomplete CQ are shown with open symbols at $t_\mathrm{CQ}=160\,\rm ms$. b) Representative RE current (linear) vs. representative RE seed (logarithmic). Blue circles - single stage H26, light green circles - staggered H26 w.~shifted deposition, dark green circles - staggered H26 w.~local deposition, black squares - single stage DTHmode24, light red squares - staggered DTHmode24 w. shifted deposition, dark red squares - staggered DTHmode24 w. local deposition,  orange diamonds - single stage DTHmode24 w.o. activated RE seed. In panel b) the seed corresponding to a single relativistic electron is marked by a vertical line; all points to the left of that line are far to the left of the scale plotted range; these points are plotted there to illustrate that the final RE current in those cases is negligible. The typical range of activated seed generated after the TQ is indicated by the gray shaded region.}
    \label{fig:correlations_current}
\end{figure*}

Fig.~\ref{fig:correlations_current} shows two correlation images, with different markers for the H26 and the DTHmode24 simulations, with single stage and staggered injection and, for the DTHmode24 scenario, with and without activated sources. In Fig.~\ref{fig:correlations_current}a we can see that there is a slight trend of reduced RE current with increasing current quench time. Such correlation is expected as a longer $t_\mathrm{CQ}$ tends to be accompanied by lower induced electric fields (during both the TQ and the CQ) to which all the runaway generation processes, in particular the Dreicer and hot-tail mechanisms, are sensitive. 

Notably, choosing the Ne content aiming to span the $t_\mathrm{CQ}=50-150\,\rm ms$ range in the baseline cases keeps $t_\mathrm{CQ}$ within or in the vicinity of the desired range also when the parameters are varied. However, we see that the H26 (blue and green circles) and the activated DTHmode24 (black and red squares) cases occupy different regions in the $I_\mathrm{RE}-t_\mathrm{CQ}$-space. The H26 cases appear at higher $t_\mathrm{CQ}$ values and span a wider range in $I_\mathrm{RE}$, extending from $\sim 12\,\rm MA$ down to negligible RE currents. The DTHmode24 cases where the activated sources are artificially removed (orange) also have negligible RE currents, while all DTHmode24 cases with activated sources included (black and red) have RE currents above $\sim 4\,\rm MA$. 

It was not possible to span the entire $50-150 \,\rm ms$ CQ time range for each scenario and injection scheme. In the H26 scenario, even a full Ne pellet was insufficient to reach the lower end, and for the DTHmode24 scenario, reaching the upper end would typically require too low Ne content, for which the TQ is incomplete (except for a couple of cases with a staggered injection and shifted deposition). In these \emph{incomplete TQ} cases a significant part of the plasma is ohmically re-heated to hundreds of $\rm eV$ following the TQ, which results in an intolerably long CQ. 

The difference between the H26 and DTHmode24 scenarios depends mostly on the difference in initial thermal energy. In H26, the relatively low thermal energy content limits the amount of material ablated before the temperature becomes too low to allow any significant further ablation (see Fig.~\ref{fig:correlations_assim} and the corresponding discussion in section \ref{sec:correlations_assim}). It is thus not possible to assimilate a sufficient amount of material to achieve a post-TQ temperature corresponding to the lower part of the desired CQ range. In contrast, in DTHmode24, the relatively high thermal energy content increases the amount of assimilated Ne, resulting in a low post-TQ temperature, while increasing the cooling required to initiate a radiative collapse. In this scenario the CQ times thus populate the lower part of the desired range, while decreasing the injected quantities in an attempt to reach the upper part often leads to an incomplete TQ.

Sometimes the ohmic current profiles resulting from an incomplete TQ are more localized, and are expected to be prone to instabilities. While worth further investigation in the future, the stability of the current channels is not monitored in our simulations. Instead, we consider all incomplete TQ cases as undesirable and we choose the Ne concentrations to avoid them. Incomplete TQs nevertheless occurred in a few cases even if the TQ was complete in the baseline cases examined when choosing the Ne concentrations; these are the cases plotted with open symbols at far right side of the scale. Note, however, that although these cases have a very long CQ time according to our definition (i.e,~measured in identical simulations without RE current), some of these actually develop significant RE currents which abort the CQ preemptively, thus their CQ time is not representative.

Negligible RE currents occur mainly because the RE seeds are extremely small in those cases. In fact, in most such cases the RE currents calculated in the simulation correspond to much less than a single electron (marked by the vertical line in Fig.~\ref{fig:correlations_current}b)\footnote{Note, that \DREAM, being a continuum code, is agnostic to the discreteness of charges, thus it can produce currents much smaller than that corresponding to a single relativistic electron. However, the cases studied here are not ambiguous in that they either produce much smaller or much larger seed currents -- corresponding either to no REs, or a sufficiently large RE population that the continuum treatment is justified. Note, that in Fig.~\ref{fig:correlations_current}b the points shown left of the "single RE in ITER" line are in fact located far to the left of this limit.}, indicating that with high probability no REs would be generated in those cases at all.

Importantly, even a tiny seed can be amplified to several MAs of final RE current. This is illustrated in Fig.~\ref{fig:correlations_current}b, showing an essentially logarithmic dependence of the final RE current on the representative seed, enabling even a seed as small as $10^{-8}\,\rm A$ to be amplified to $\sim 5\,\rm MA$. To lead the eye, the trends found in the DTHmode24 and H26 cases are shown with dashed lines, and while the values and the slopes are comparable between the two plasmas, we find that the same seed typically yields a higher final RE current in the H26 plasma. This must be caused by a higher avalanche rate, which is consistent with the typically poorer hydrogen assimilation in H26 caused by the lower initial temperature (see Fig.~\ref{fig:correlations_assim}), which leads to lower effective critical electric fields. In addition, while the H26 cases extend to very small seeds (and even to zero seed REs, outside the scale of Fig.~\ref{fig:correlations_current}b), in the DTHmode24 cases the seed cannot be reduced below the minimum representative Compton RE seed of $\sim 10^{-3}\,\rm A$. As a result, we find no RE currents below $\sim 4\,\rm MA$ in DTHmode24.

\subsubsection{Assimilation}
\label{sec:correlations_assim}

\begin{figure*}[!h]
    \begin{minipage}[h]{0.79\textwidth}
    \includegraphics[width=0.5\textwidth]{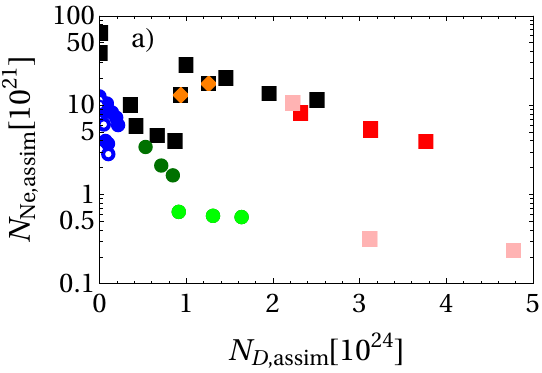}
    \includegraphics[width=0.5\textwidth]{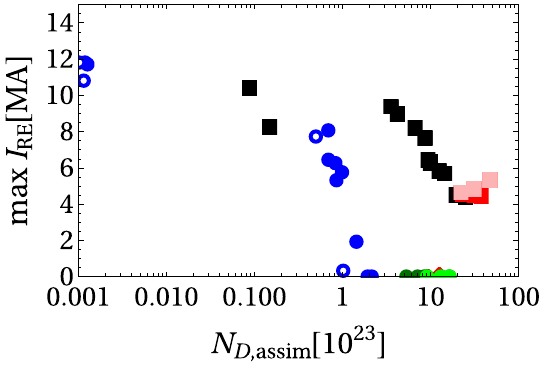}
    \put(-155,115){\textcolor{black}{b)}}
    \end{minipage}
    \begin{minipage}[h]{0.2\textwidth}
    \includegraphics[width=\textwidth]{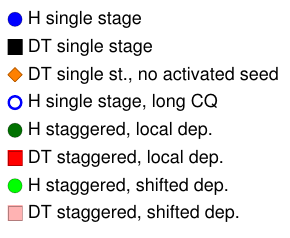}
        \end{minipage}
    \caption{Correlations between figures of merit involving assimilation of the injected material. a)  Number of assimilated Ne (logarithmic) vs. D atoms (linear). b) Representative RE current (linear) vs. number of assimilated D atoms (logarithmic). Blue circles - single stage H26, light green circles - staggered H26 w.~shifted deposition, dark green circles - staggered H26 w.~local deposition, black squares - single stage DTHmode24, light red squares - staggered DTHmode24 w. shifted deposition, dark red squares - staggered DTHmode24 w. local deposition,  orange diamonds - single stage DTHmode24 w.o. activated RE seed.}
    \label{fig:correlations_assim}
\end{figure*}

The lower CQ times seen in figure \ref{fig:correlations_current}a for the DTHmode24 cases can be linked to higher ablation figures caused by the significantly higher initial thermal energy, as mentioned in section \ref{sec:correlations_current}. This is reflected in the assimilated Ne and D quantities shown in  Fig.~\ref{fig:correlations_assim}a, where the DTHmode24 cases typically occupy higher values in both Ne and D than H26 (with an exception for a couple of the staggered DTHmode24 cases with low Ne assimilation).

We also see that there is a negative, approximately exponential, correlation between the assimilated Ne and D, with different slopes for different groups of points.  The negative trend occurs because a smaller assimilated Ne quantity reduces the radiative cooling, so that the temperature remains higher in the vicinity of the shards, allowing for a faster ablation and thus a higher assimilated D quantity. Conversely, a larger amount of D gives a stronger dilution cooling, which slows down the ablation and thus reduces the amount of assimilated Ne. 

For the single stage injections (blue for H26, black and orange for DTHmode24), the different slopes correspond to the cases with early and late TQ. For the staggered cases (which were only performed with a late TQ), the division occurs between the cases with and without the shift of the deposition included. Another clear trend is that the staggered cases typically have larger amounts of assimilated D and, consequently, lower amounts of assimilated Ne.

The RE current is rather strongly negatively correlated with the amount of assimilated D, as shown in Fig.~\ref{fig:correlations_assim}b. This is partly due to the negative correlation between the amount of assimilated Ne and D, as smaller Ne quantities lead to later TQ onsets, longer separation between the cooling phases and lower electric fields, all of which reduce the RE generation. A higher D quantity also has the direct effect of increasing the collisional damping. As a result, both the staggered H26 cases and the DTHmode24 cases without activated sources all have negligible RE currents. There are also two single stage H26 cases with negligible RE currents, one of which has the lowest amount of injected Ne among the H26 cases ($N_\mathrm{Ne,inj}=5\cdot10^{22}$), and one of which uses three pellets. Both of them used TQ conditions favourable for RE avoidance, i.e.~a late TQ and $t_\mathrm{TQ}=3\,\rm ms$, leading to a relatively large ablation, long separation between the cooling phases and small hot-tail seed. 

Increasing the number of shards (i.e. using the small shard size) can also increase the D assimilation, but the effect is rather moderate ($\sim 10\%$). Moreover, it also increases the assimilated Ne, which can even lead to a moderate increase in the RE current. However, the staggered cases appear to more robustly suppress the RE formation. 

At the highest assimilated D quantities obtained with the DTHmode24 scenario, the decreasing trend of $I_\mathrm{RE}$ with $N_\mathrm{D,assim}$ flattens out, and a hint of an emerging increasing trend can be seen. The reason for this is that large enough D quantities can make the hydrogen species recombine, while there is still a significant ohmic current left in the plasma that can be converted to REs. At high electric fields, this decreases the collisional damping while retaining the number of available targets for the RE avalanche, resulting in an increase in the RE current, as discussed in \cite{Vallhagen2020Runaway}.

\subsection{Cases with reduced plasma current}
\label{sec:reduced_Ip}

We now turn our attention to the scenarios with lower initial plasma current (He56 and H123). As the maximum avalanche gain scales exponentially with the plasma current, we now expect significantly smaller conversion rates for a given seed compared to the $15\,\rm MA$ scenarios studied above. This is indeed confirmed by the simulation results, although it still seems important to use a staggered injection to robustly avoid a significant RE formation for all realistic TQ conditions.

\begin{figure}
    \centering
    \includegraphics[trim={0 1.15cm 0 0},clip,width=\columnwidth]{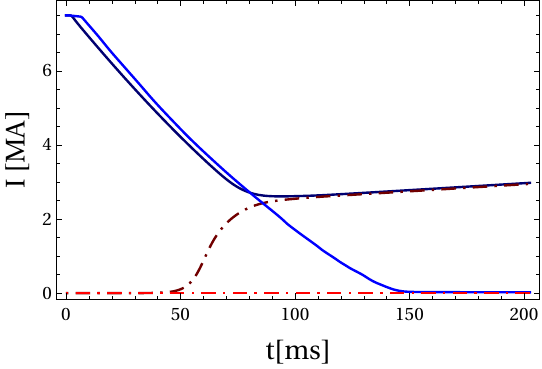}
    \put(-190,120){a)}
    \put(-90,120){\footnotesize He56}
    \put(-90,110){\footnotesize $N_\mathrm{Ne} = 5\cdot 10^{21}$}
    \put(-90,100){\footnotesize Early TQ, $t_\mathrm{TQ}= 1$ ms}
    \put(-100,55){\footnotesize Single stage}
    \put(-75,15){\footnotesize Staggered}
    
    \includegraphics[width=\columnwidth]{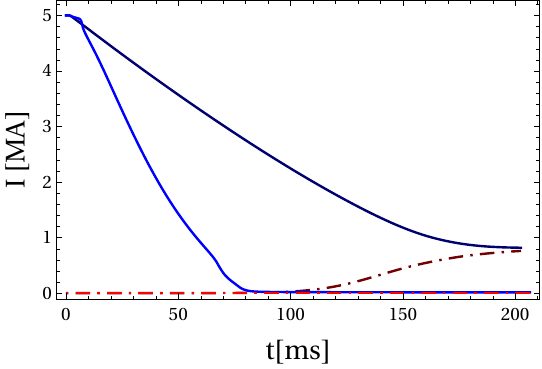}
    \put(-190,150){b)}
    \put(-90,150){\footnotesize H123}
    \put(-90,140){\footnotesize Early TQ, $t_\mathrm{TQ}= 1$ ms}
    \put(-115,105){\footnotesize Single stage}
    \put(-115,95){\footnotesize $N_\mathrm{Ne} = 10^{21} $, 3 pellets}
    \put(-150,70){\footnotesize Staggered}
    \put(-150,60){\footnotesize $N_\mathrm{Ne} = 1.83\cdot 10^{24}$}
    
    \caption{Current evolutions following a disruption with unfavourable TQ conditions mitigated with a single stage injection and a staggered injection, for a) the He56 scenario and b) the H123 scenario. For the He56 scenario, the injected Ne quantity is $N_\mathrm{Ne} = 5\cdot 10^{21}$ atoms with both the single stage and the staggered injection. For the H123 scenario, the injected Ne quantity is $N_\mathrm{Ne} = 1.83\cdot 10^{24}$ atoms for the staggered injection, and $N_\mathrm{Ne} = 10^{21} $ atoms for the single stage injection, in the latter case injected with 3 pellets.}
    \label{fig:reduced_Ip}
\end{figure}

Figure \ref{fig:reduced_Ip} shows representative current evolutions following a single stage and a staggered injection with unfavourable (early and short) TQ conditions, for a) the He56 scenario and b) the H123 scenario. For the He56 scenario, we inject $N_\mathrm{Ne} = 5\cdot 10^{21}$ Ne atoms for both the single stage and staggered injection. 
In the staggered injection case with the H123 scenario, we inject $N_\mathrm{Ne} = 1.83\cdot 10^{24}$ Ne atoms.
For the single stage injection, we decrease the Ne quantity and increase the D quantity to assess whether it is feasible to avoid a significant RE formation even with a single stage injection if one would be allowed to exceed the upper CQ time limit. In the case displayed we use 3 pellets with a total of $N_\mathrm{Ne} = 10^{21} $ Ne atoms.

We see in Fig.~\ref{fig:reduced_Ip}a) that a disruption with unfavourable TQ conditions may result in several MAs of RE current even with a plasma current of $7.5\,\rm MA$, if it is only mitigated by a single stage injection; in this case the representative RE current is $2.48\,\rm MA$. 
Similarly, in figure \ref{fig:reduced_Ip}b) we see that the RE current could not be reduced much below $1\,\rm MA$ with a single stage injection even with a $5\,\rm MA$ initial plasma current, despite exceeding the upper CQ time limit; the displayed case has a representative RE current of $0.77\,\rm MA$. 

The representative RE seed is close to $3\,\rm A$ in both single stage injection cases, demonstrating the potential severity of the RE avalanche also at these reduced plasma currents in ITER. The mitigation is also somewhat impeded by a rather low assimilation ($\sim 10\%$) for the single stage injections. This is similar to the corresponding cases with the H26 scenario, despite the $\sim 2$ times higher temperature, as the density is significantly lower, making the plasma easier to cool.

In the staggered cases, however, we see no significant RE formation, regardless of the Ne quantity in the second stage (as long as the CQ remains complete). The same applies to single stage injections with favourable TQ conditions, except for extreme cases with very large Ne quantities. We also note that the RE conversion is considerably smaller for a given seed than in the $15\,\rm MA$ scenarios; at a representative seed of $3\,\rm A$, we find a conversion of about $50\%$ for the $15\,\rm MA$ scenarios (see figure \ref{fig:correlations_current}b), while the single stage injection cases in figure \ref{fig:reduced_Ip} have conversions of $33\%$ for the He56 scenario and $15\%$ for the H123 scenario.

\subsection{Cases with injection arriving after the TQ}
\label{sec:reheat}

\begin{figure}
	\centering
    \includegraphics[width=\columnwidth]{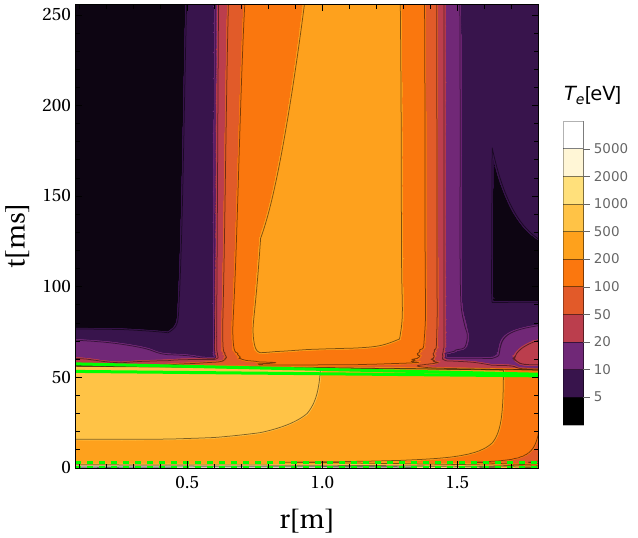}
	\put(-200,40){\textcolor{black}{a)}}
    \put(-150,40){\textcolor{black}{\tiny $\delta B/B=4\cdot 10^{-4}$}}
    
	\includegraphics[width=\columnwidth]{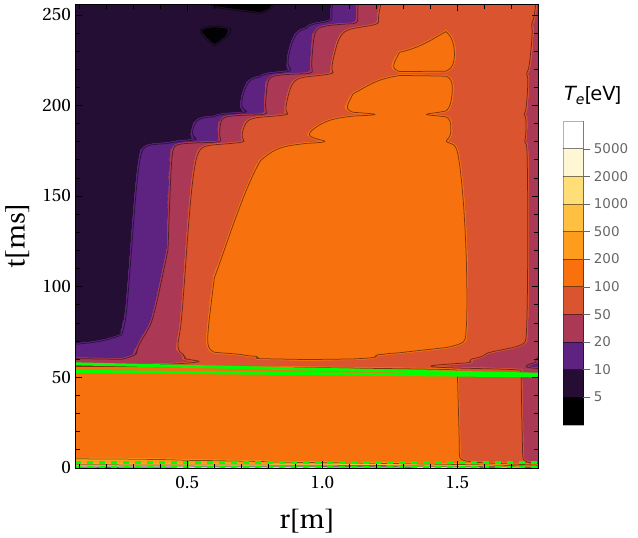}
	\put(-200,40){\textcolor{black}{b)}}
    \put(-160,40){\textcolor{black}{\tiny $\delta B/B=2\cdot 10^{-3}$}}
	\caption{\label{fig:lateinj} Late injection case based on the H26 scenario, using $99\%$ neon pellets injected $50\,\rm ms$ after the end of the transport event that started at $t=0$. The cases employ $t_{\rm TQ}=3\,\rm ms$ and a remnant electron heat transport corresponding to a) $\delta B/B=4\cdot 10^{-4}$, and b) $\delta B/B=2\cdot 10^{-3}$.}
\end{figure}

Finally we consider cases where the injection is triggered too late, so that it arrives after the TQ transport event; specifically after a $50\,\rm ms$ delay. All the late injection cases with standard injection settings lead to re-heating and incomplete current quench, as illustrated in Fig.~\ref{fig:lateinj}, which is based on the H26 scenario. Here we use conditions that would favor avoiding a re-heating, namely inject a $99\%$ neon pellet and employ a remnant electron heat transport after the TQ corresponding to $\delta B/B=4\cdot 10^{-4}$ and $2\cdot 10^{-3}$ (note that the latter is comparable with $\delta B /B$ during a TQ). We see that after the TQ, the temperature increases due to ohmic heating to $T_\mathrm{e}\lesssim 1\,\rm keV$ with a remnant $\delta B/B=4\cdot 10^{-4}$, and $T_\mathrm{e}\lesssim 200\,\rm eV$ with a remnant $\delta B/B=2\cdot 10^{-3}$. When the injection arrives, the temperature drops to the $10-100\,\rm eV$ range, but a large part of the plasma re-heats again to a temperature of a few hundred eV. In other cases we used a fixed remnant heat diffusivity of $1\,\rm m^2/s$, and obtained qualitatively similar results.

The reason for the re-heating is the extremely low assimilation rate of the pellet, owing to the low plasma temperature ($<1 \,\rm keV$). Thus, increasing the remnant heat diffusivity does not help, as it just decreases the temperature further, and along with it the assimilation rate (it is $0.4\%$ for $\delta B/B=4\cdot 10^{-4}$ and $0.2\%$ for $\delta B/B=2\cdot 10^{-3}$). Similarly, moving the time of the injection closer to the transport event, when the plasma temperature has had less time to re-grow, further aggravates the re-heating problem. In conclusion, it is very important for the utility of SPI that the pellets do not arrive late with respect to the TQ. 

The re-heating can however, in some cases, be counteracted by reducing the shard size, and thereby increasing the assimilation. This was possible in an He56 scenario with $t_\mathrm{TQ}=3\,\rm ms$, a 99\% Ne pellet injection and a fixed remnant heat diffusivity of $1\,\rm m^2/s$, where we increased the number of shards from $487$ to $5185$ (corresponding to the \emph{standard} and \emph{small} shard sizes as named in \cite{AkinobuITPA}). This increased the assimilation rate from $0.19\%$ to $0.31\%$ that, in this case, was sufficient to eliminate the re-heating issue. Nevertheless, reducing the shard size to the range considered here does not appear to be a robust method to avoid re-heating, although the assimilation might be further enhanced by employing a more extreme shattering. To ameliorate this problem ITER plans to employ three upper injectors dedicated to post-TQ injection, aiming at injecting mainly gas by using a very large shattering angle. Assessing gas penetration is however out of scope of the present study.

It is also important to mention that the case where re-heating was avoided produced an almost full runaway conversion; as no injected material had reached the plasma at the time of the TQ, the collisional drag remained rather low, making the plasma very prone to RE seed generation. Here we have however assumed the TQ duration to remain in the $1$--$3\,\rm ms$ considered for mitigated disruptions, while the TQ duration in ITER in the absence of impurities might be substantially longer, up to several tens of milliseconds \cite{Strauss_2023}. This could significantly reduce the RE seed generation, and also expand the time range during which the plasma is hot enough to significantly ablate the pellet, allowing for a later pellet arrival.

\section{Discussion}
\label{sec:discussion}
The results presented here are based on modeling choices which are, in many ways, conservative, and it is thus possible that, in particular, the RE currents calculated here are overestimated. One reason for this is that \DREAM{} does not account for the vertical motion of the plasma and the resulting shrinking of the closed field line region. As the REs will very quickly get lost to the wall on the open field lines, the vertical motion acts to decrease the volume with unhindered runaway formation.

Although a detailed study of the effect of a vertical displacement event is outside the scope of the paper, one can get an indication of the impact of this effect by varying other aspects of the boundary condition at the conducting vessel wall, with the aim to emulate the evolution of the poloidal flux available for RE generation on a given flux surface. In particular, setting a perfectly conducting wall very close to the plasma, thus forcing the poloidal flux at the edge to remain constant, could reduce the poloidal flux variation and the resulting avalanche gain considerably. Such a boundary condition would also make the simulated poloidal flux variation on a given flux surface similar to the actual variation during the time the flux surface is closed, if the plasma moves in such a way that the poloidal flux at the last closed flux surface remains approximately constant. This is expected to be the case if the resistive time scale of the plasma wall is long compared to the CQ time, which will be true for mitigated disruptions in ITER, and is also observed in recent ITER simulations with the JOREK code \cite{WangREM}.

\begin{figure}
    \centering
       \includegraphics[width=1\columnwidth]{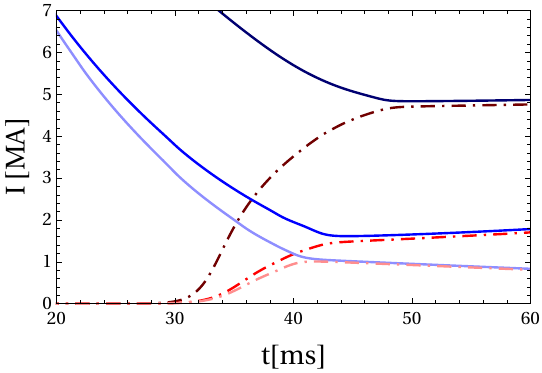}
    \put(-115,140){\footnotesize $r_\mathrm{wall}=2.83$ m, $t_\mathrm{wall}=0.5$ s}
    \put(-115,70){\footnotesize $r_\mathrm{wall}=r_\mathrm{minor}$, $t_\mathrm{wall}=0.5$ s}
    \put(-105,40){\footnotesize $r_\mathrm{wall}=r_\mathrm{minor}$, $t_\mathrm{wall}=\infty$}
    \caption{\label{fig:IpWallEffect} Time evolution of the plasma current and RE current using different settings for the wall boundary condition, for the staggered DTHmode24 case using 1 pure D pellet  for the first injection and a pellet with $1.35\%$ Ne for the second injection (target CQ time 100 ms for the corresponding baseline case).}    
\end{figure}

To illustrate the range of possible outcomes in case the poloidal flux available for runaway generation is modified by some process, figure \ref{fig:IpWallEffect} shows a comparison between the current evolutions obtained with different wall boundary conditions. The baseline case with an effective conducting wall radius of $r_\mathrm{wall}=2.833\,\rm m $ and resistive wall time of $t_\mathrm{wall} = 500\,\rm ms $ is compared to a case with $r_\mathrm{wall}=r_\mathrm{minor}$ (where $r_\mathrm{minor}$ is the plasma minor radius) and $t_\mathrm{wall} = 500\,\rm ms$, and one with $r_\mathrm{wall}=r_\mathrm{minor}$ and $t_\mathrm{wall} = \infty$. The comparison is made for the D-T case with the lowest RE current included in figure \ref{fig:correlations_assim}b (excluding the cases without the shifted deposition for pure D injections). This case uses a staggered injection with 1 pure D pellet for the first injection and a Ne concentration of $1.35\% $ in the second injection. 

Reducing the wall radius to $r_\mathrm{wall}=r_\mathrm{minor}$ is found to reduce the RE current by several MAs. When also setting $t_\mathrm{wall} = \infty$, the RE current becomes as low as $\sim 1\,\rm MA$. While using $t_\mathrm{wall} = 500\,\rm ms$ increases the RE current to $\sim 2\,\rm MA$ compared to the $t_\mathrm{wall} = \infty$ case. The RE current also keeps increasing during the plateau phase as additional poloidal flux can diffuse through the wall. Conversely, with $t_\mathrm{wall} = \infty$, a clear RE decay is observed. It may be however, that the behavior observed here during the plateau phase is not realized in an experiment, since the flux surfaces may completely disintegrate before the RE plateau is fully formed. 

It should be noted that the reduction might not be quite as large in reality; for instance, the magnetic energy carried by the escaping REs will not immediately be lost from the system, but will remain in the form of a halo current, and the electric field associated with this halo current might diffuse into the confined plasma region where it can contribute to the RE generation. Nevertheless, this exercise indicates that the effect of the plasma motion, and in general modified wall boundary conditions, may result in a major reduction in the generated RE current, motivating more detailed future studies on this front.

An additional conservative modelling choice is that no RE transport losses due to magnetic perturbations are included during the CQ. While the overall perturbation level is expected to be much smaller than during the TQ, a $\delta B/B$ as low as $\sim 10^{-4}$ may be sufficient to have a notable impact on the RE generation and decay \cite{Svensson2021Effects}. An even larger uncertainty stems from the unknown time of reformation of the outermost flux surfaces after the TQ. The RE current may also be reduced when accounted for by a kinetic calculation, as the fluid RE generation rates, in particular for hot-tail generation, have been shown to overestimate the RE generation \cite{EkmarkMSc}. 

There are also modelling choices which may lead to an underestimation of the RE generation, although perhaps less significant compared to the overestimation effects mentioned above. For instance, in this study we have not accounted for the rapid current density profile relaxation and the corresponding current spike typically taking place during the TQ. This effect has been shown to increase the RE generation in cases prone to RE generation in the outer part of the plasma \cite{IstvanHyperres}. However, a shift of the RE current density profile towards the edge would also make the REs more prone to be lost to the wall during a vertical displacement event; the result of the combination of these competing effects remains to be investigated.

Another uncertainty regarding the RE generation is that the $\gamma$-photon flux and spectrum was calculated for a first wall made of beryllium, while ITER currently plans to have a first wall made of tungsten. This might alter the $\gamma$-photon flux, and consequently the Compton scattering RE seed generation. However, as observed in figure \ref{fig:correlations_current}b), the final RE current (if macroscopic) is only logarithmically sensitive to the RE seed; approximately $0.4$--$0.9\,\rm MA$ change in the final RE current per order of magnitude change in the seed. Thus the $\gamma$-photon flux must change by orders of magnitude before it would lead to a major difference in the final RE current. Moreover, a recent assessment shows that the difference in the prompt $\gamma$-photon flux and spectrum between a tungsten and a beryllium first wall is not significant \cite{Eduardo}. While the corresponding differences in the decay $\gamma$-photon flux and spectrum present during the CQ remain to be assessed, current estimates indicate that the change of the first wall material is not likely to have a major impact on the RE generation.

Finally, there are several sources of uncertainty which are not directly related to the RE generation. One of these concerns the details of the transport event. We have attempted to cover the plausible range of onset criteria and magnetic perturbation levels, but recent calculations indicate that the TQ time scale 
can be longer than assumed here. This could reduce the hot-tail seed considerably, possibly allowing for a complete RE avoidance in a wider range of scenarios with a single stage injection without activated RE sources. Moreover, the ion transport coefficients are a significant source of uncertainty. While our values are successfully chosen to match the mixing time scale found in 3D MHD simulations, this argument is not sufficient to accurately constrain the involved parameters, and thus both the time scale of the ion mixing and the resulting radial profile might vary considerably. This could have a particular importance for the staggered injection cases with a shifted deposition, which are dependent on the ion mixing for material deposition in the core. In addition, it is worth emphasising that the shift of the deposition has here been assumed to be moderate, while there might occur shifts comparable to the plasma minor radius \cite{Vallhagen2023Drift}, which could eject substantial amounts of material from the plasma and thereby reduce the assimilation.

\section{Conclusion}
\label{sec:conclusion}
We have explored a wide range of feasible disruption mitigation SPI settings and conditions for four ITER scenarios with a focus on RE avoidance: one L-mode hydrogen plasma with a $15\,\rm MA$ plasma current (H26); one H-mode helium plasma with a $7.5\,\rm MA$ plasma current (He56); one H-mode hydrogen plasma with a $5\,\rm MA$ plasma current (H123), all without activation of the wall; and one DT H-mode scenario with a $15\,\rm MA$ plasma current, including activated runaways sources (DTHmode24). The disruption mitigation performance was evaluated based on the CQ time, which should be in a given range -- $50-150\,\rm ms$ in $15\,\rm MA$ discharges -- and the RE current, which should be lower than $150\,\rm kA$. We observe a wide range of runaway conversions from $80\%$ at the largest Ne and lowest D quantities and unfavourable TQ conditions, down to negligible values for low Ne quantities, large D quantities and favourable TQ conditions.

The most important conclusions that can be drawn from this study are the following:
\begin{enumerate}
   \item The two-stage injection is an effective means to suppress the runaway current. It suppresses the hot-tail seed generation by leaving time for the tail of the distribution to thermalize, and it reduces the avalanche rate by improving hydrogen assimilation. In particular, with two-stage injection the runaway current can be completely eliminated in non-activated discharges even at $15\,\rm MA$ initial plasma current.
   \item While two-stage injection also yields the best performing cases in activated scenarios, there the irreducible nuclear seeds -- active even after the transport losses of the TQ -- yield a floor of $\approx 4.5 \,\rm MA$ in the final RE current (corresponding to a $\rm mA$-level seed).
   \item Even in the reduced initial current scenarios it is possible to get multi-$\rm MA$ runaway currents, especially for unfavorable (early and short) TQ conditions. However, in these cases the two-stage injection can robustly eliminate the runaway current. 
   \item Both concerning $t_{\rm CQ}$ and the final RE current, it is important that the pellet injections arrive before the TQ. Otherwise, the poor ablation in the much cooler background plasma leads to insufficient material assimilation and a corresponding ohmic re-heating of the plasma, along with an intolerably long CQ time. Although the poor assimilation may be ameliorated by post-TQ gas injection and extreme pellet shattering, this will likely not be sufficient to compensate for the lack of collisional damping of the RE seed generation during the TQ. Thus, unexpectedly high runaway currents may be generated.
   \item The current quench time can be kept within the desired range in all plasma scenarios considered, with appropriate injected quantities.
   \item The final RE current, if it is macroscopically large, is logarithmically sensitive to the effective runaway seed current (i.e., the RE current surviving the TQ losses, plus the seed generated afterwards). Megaampere-scale  currents may arise from a single runaway electron in the avalanche-dominated regime of ITER. 
   \item The external electric field boundary conditions have a major effect on the final runaway current. This includes the location and the resistive timescale of the wall, as well as effective changes in these boundary conditions that arise due to a scraping off of the current channel during vertical displacement events. 
   \item A negative correlation with an exponential sensitivity of the assimilated neon quantity to the assimilated hydrogen quantity is observed. Staggering tends to yield higher hydrogen and smaller assimilated neon quantities. The overall assimilation figures are significantly higher in the hot DTHmode24 scenario than in the H26 scenario. 
   \item A high assimilated hydrogen content is  beneficial in most cases; it is a requisite to obtain scenarios with negligible runaway current. However at the highest assimilated values (few times $10^{24}$ assimilated atoms), which are achievable only in DTHmode24, we observe the previously predicted increase of runaway current caused by hydrogen recombination.  
\end{enumerate}

The results presented here are based on conservative modelling choices, and they should prompt continued experimental validation of \DREAM, as well as further studies accounting for additional effects. For instance the available poloidal magnetic flux can potentially be reduced, and runaway losses enhanced, during a vertical displacement event. In addition, the possibility of RE instabilities and remnant magnetic perturbations after the thermal quench may also reduce the RE current. 

\section*{Acknowledgement}
The authors are grateful to A Matsuyama, D Hu, I Ekmark, L Antonsson and P Helander for fruitful discussion. This work has been carried out in collaboration with the ITER Organisation under implementing agreement IO/IA/21/4300002402. ITER is the Nuclear Facility INB No.~174. This paper explores physics processes during the plasma operation of the tokamak when disruptions take place; nevertheless the nuclear operator is not constrained by the results presented here. The views and opinions expressed herein do not necessarily reflect those of the ITER Organization. The work has been partly carried out within the framework of the EUROfusion Consortium, funded by the European Union via the Euratom Research and Training Programme (Grant Agreement No 101052200 — EUROfusion). Views and opinions expressed are however those of the authors only and do not necessarily reflect those of the European Union or the European Commission. Neither the European Union nor the European Commission can be held responsible for them.

\bibliography{bibliography.bib}

\providecommand{\newblock}{}
\begin{thebibliography}{10}
\expandafter\ifx\csname url\endcsname\relax
  \def\url#1{{\tt #1}}\fi
\expandafter\ifx\csname urlprefix\endcsname\relax\def\urlprefix{URL }\fi
\providecommand{\eprint}[2][]{\url{#2}}

\bibitem{LehnenITER}
Lehnen M and {ITER Disruption Mitigation Task Force} 2021 The {ITER}
  {Disruption} {Mitigation} {System} -- {Design} progress and design validation
  Theory and Simulation of Disruptions Workshop, online
  \urlprefix\url{https://tsdw.pppl.gov/Talks/2021/Lehnen.pdf}

\bibitem{Martin2017Formation}
Mart{\' i}n-Sol{\' i}s J, Loarte A and Lehnen M 2017 {\em Nuclear Fusion\/}
  {\bf 57} 066025

\bibitem{Vallhagen2020Runaway}
Vallhagen O, Embreus O, Pusztai I, Hesslow L and F{\" u}l{\" o}p T 2020 {\em
  Journal of Plasma Physics\/} {\bf 86} 905890204

\bibitem{Pusztai2023Bayesian}
Pusztai I, Ekmark I, Bergstr{\" o}m H, Halldestam P, Jansson P, Hoppe M,
  Vallhagen O and F{\" u}l{\" o}p T 2023 {\em Journal of Plasma Physics\/} {\bf
  89} 905890204

\bibitem{Vallhagen2022Effect}
Vallhagen O, Pusztai I, Hoppe M, Newton S and F{\" u}l{\" o}p T 2022 {\em
  Nuclear Fusion\/} {\bf 62} 112004

\bibitem{Nardon2020Fast}
Nardon E, Hu D, Hoelzl M and Bonfiglio D 2020 {\em Nuclear Fusion\/} {\bf 60}
  126040

\bibitem{Hoppe2021DREAM}
Hoppe M, Embreus O and F{\" u}l{\" o}p T 2021 {\em Computer Physics
  Communications\/} {\bf 268} 108098

\bibitem{Kim_2018}
Kim S, Casper T and Snipes J 2018 {\em Nuclear Fusion\/} {\bf 58} 056013
  \urlprefix\url{https://dx.doi.org/10.1088/1741-4326/aab034}

\bibitem{Huysmans2007MHD}
Huysmans G and Czarny O 2007 {\em Nuclear Fusion\/} {\bf 47} 659--666

\bibitem{Samulyak2021Lagrangian}
Samulyak R, Yuan S, Naitlho N and Parks P 2021 {\em Nuclear Fusion\/} {\bf 61}
  046007

\bibitem{Parks2016Modeling}
Parks P 2016 Modeling {Dynamic} {Fracture} of {Cryogenic} {Pellets} Tech. rep.
  {Office of Scientific and Technical Information (OSTI)}

\bibitem{AkinobuITPA}
Matsuyama A 2023 Integrated {1.5D SPI} modelling with {INDEX} 42nd meeting of
  ITPA MHD Disruption and Control Topical Group

\bibitem{Parks2000Radial}
Parks P~B, Sessions W~D and Baylor L~R 2000 {\em Physics of Plasmas\/} {\bf 7}
  1968--1975

\bibitem{Rozhansky2004Mass}
Rozhansky V, Senichenkov I, Veselova I and Schneider R 2004 {\em Plasma Physics
  and Controlled Fusion\/} {\bf 46} 575--591

\bibitem{Pegourie2006Homogenization}
P{\' e}gouri{\' e} B, Waller V, Nehme H, Garzotti L and G{\' e}raud A 2006 {\em
  Nuclear Fusion\/} {\bf 47} 44--56

\bibitem{Vallhagen2023Drift}
Vallhagen O, Pusztai I, Helander P, Newton S and F{\" u}l{\" o}p T 2023 {\em
  Journal of Plasma Physics\/} {\bf 89} 905890306

\bibitem{ADAS}
Summers H~P 2004 The {ADAS} user manual, version 2.6
  \url{http://www.adas.ac.uk}

\bibitem{loarteTQ}
Loarte A, Andrew P, Matthews G~F, Paley J, Riccardo V, Counsell G, Eich T,
  Fuchs C, Gruber O, Herrmann A, Pautasso G, Federici G, Finken K~H, Maddaluno
  G, Whyte D and {DIII-D National Fusion Facility, San Diego, CA (United
  States)} 2005 Expected energy fluxes onto iter plasma facing components
  during disruption thermal quenches from multi-machine data comparisons Proc.
  20th Int. Conf. Vilamoura, 2004
  \urlprefix\url{https://www.osti.gov/etdeweb/biblio/20641491}

\bibitem{Rechester1978Electron}
Rechester A~B and Rosenbluth M~N 1978 {\em Physical Review Letters\/} {\bf 40}
  38--41

\bibitem{Svensson2021Effects}
Svensson P, Embreus O, Newton S~L, S{\" a}rkim{\" a}ki K, Vallhagen O and F{\"
  u}l{\" o}p T 2021 {\em Journal of Plasma Physics\/} {\bf 87} 905870207

\bibitem{Hu2021Radiation}
Hu D, Nardon E, Hoelzl M, Wieschollek F, Lehnen M, Huijsmans G, van Vugt D~C,
  Kim S~H, contributors J and team J 2021 {\em Nuclear Fusion\/} {\bf 61}
  026015

\bibitem{Linder2020Self}
Linder O, Fable E, Jenko F, Papp G and Pautasso G 2020 {\em Nuclear Fusion\/}
  {\bf 60} 096031

\bibitem{NN_Dreicer}
Hesslow L, Unnerfelt L, Vallhagen O, Embreus O, Hoppe M, Papp G and Fülöp T
  2019 {\em Journal of Plasma Physics\/} {\bf 85} 475850601

\bibitem{Hesslow_fluid_ava}
Hesslow L, Embr{\'{e}}us O, Vallhagen O and Fülöp T 2019 {\em Nuclear
  Fusion\/} {\bf 59} 084004

\bibitem{Matsuyama_2022}
Matsuyama A, Hu D, Lehnen M, Nardon E and Artola J 2022 {\em Plasma Physics and
  Controlled Fusion\/} {\bf 64} 105018
  \urlprefix\url{https://dx.doi.org/10.1088/1361-6587/ac89b2}

\bibitem{Strauss_2023}
Strauss H~R, Chapman B~E and Hurst N~C 2023 {\em Plasma Physics and Controlled
  Fusion\/} {\bf 65} 084002
  \urlprefix\url{https://dx.doi.org/10.1088/1361-6587/acdff8}

\bibitem{WangREM}
Wang C 2023 Simulation of runaway electron generation during disruptions with
  vertical displacement in {ITER} using {JOREK} 10th Runaway Electron Modelling
  (REM) meeting, Max Planck Institute for Plasma Physics, Garching, Germany
  \urlprefix\url{https://ft.nephy.chalmers.se/?p=abstract\&id=69}

\bibitem{EkmarkMSc}
Ekmark I 2023 {\em Multi-objective Bayesian optimization of tokamak disruptions
  using fluid and kinetic models\/} Master's thesis Chalmers University of
  Technology \urlprefix\url{http://hdl.handle.net/20.500.12380/306024}

\bibitem{IstvanHyperres}
Pusztai I, Hoppe M and Vallhagen O 2022 {\em Journal of Plasma Physics\/} {\bf
  88} 905880409

\bibitem{Eduardo}
Polunovskiy E 2023 Private communication

\end{thebibliography}
\appendix
\section{Magnetic perturbation amplitudes corresponding to different TQ times}
\label{app:dBB}
In table \ref{tab:dBB} we list the values of the magnetic perturbation amplitude $dB/B$ found to give a TQ time of 1 or 3 ms, with transport as the only loss channel, for all four studied scenarios.
\begin{table}[h]
    \centering
    \begin{tabular}{c|c|c}
       Scenario & $t_\mathrm{TQ}\,\rm [ms]$ & $\delta B/B\,\rm [\%]$ \\\hline
        H26 & 1 & 0.374 \\
        H26 & 3 & 0.216 \\\hline
        DTHmode24 & 1 & 0.348 \\
        DTHmode24 & 3 & 0.201 \\\hline
        He56 & 1 & 0.350\\
        He56 & 3 & 0.202\\\hline
        H123 & 1 & 0.344\\
        H123 & 3 & 0.199\\
    \end{tabular}
    \caption{List of magnetic perturbation amplitude $dB/B$ corresponding to a TQ time of 1 or 3 ms, for all four scenarios studied in this paper.}
    \label{tab:dBB}
\end{table}

\section{List of simulations at full plasma current}
\label{app:settings}
In table \ref{tab:settings} we list the settings for the simulations performed with a $15\,\rm MA$ plasma current with injection before the TQ, which are included in Figs.~\ref{fig:correlations_current}-\ref{fig:correlations_assim}. The number of pellets injected in the first and second stage are denoted by $N_\mathrm{p1}$ and $N_\mathrm{p2}$, respectively, and the other settings are defined in section \ref{sec:settings}. The settings not included in the table are identical for all simulations, and are also specified in section \ref{sec:settings}.

\begin{table*}[h]
    \centering
    \begin{tabular}{c|c|c|c|c|c|c|c|c|c|c}
\# & Scenario & TQ & $t_\mathrm{TQ}$& $N_\mathrm{p1}$ & $N_\mathrm{p2}$ & $N_\mathrm{Ne}$ & Shard size & Deposition & Act. & Comment\\
        & & onset & [ms]& & & [$10^{22}$] & & & REs? & \\\hline
        S1 & H26 & Late & 3 & 1 & 0 & $183$& Default & local & No & Favourable\\
        S2 & H26 & Late & 3 & 1 & 0 & $20$& Default & local & No & TQ for RE\\
        S3 & H26 & Late & 3 & 1 & 0 & $5.22$& Default & local & No & avoidance\\\hline
        S4 & H26 & Early & 1 & 1 & 0 & $183$& Default & local & No & Unfavourable\\
        S5 & H26 & Early & 1 & 1 & 0 & $20$& Default & local & No & TQ for RE\\ 
        S6 & H26 & Early & 1 & 1 & 0 & $5.22$& Default & local & No & avoidance\\\hline
        S7 & H26 & late & 1 & 1 & 0 & $183$& Default & local & No & Intermediate\\
        S8 & H26 & Early & 3 & 1 & 0 & $183$& Default & local & No & TQ\\\hline
        S9 & H26 & Late & 3 & 1 & 0 & $20$& Large & local & No & Varying\\
        S10 & H26 & Late & 3 & 1 & 0 & $20$& Small & local & No & shard size\\\hline
        S11 & DTHmode24 & Late & 3 & 1 & 0 & $150$& Default & local & Yes & Favourable\\
        S12 & DTHmode24 & Late & 3 & 1 & 0 & $2.5$& Default & local & Yes & TQ for RE\\
        S13 & DTHmode24 & Late & 3 & 1 & 0 & $5.22$& Default & local & Yes & avoidance \\\hline
        S14 & DTHmode24 & Early & 1 & 1 & 0 & $150$& Default & local & Yes &Unfavourable \\
        S15 & DTHmode24 & Early & 1 & 1 & 0 & $2.5$& Default & local & Yes & TQ for RE\\
        S16 & DTHmode24 & Early & 1 & 1 & 0 & $5.22$& Default & local & Yes & avoidance \\ \hline
        S17 & DTHmode24 & Late & 3 & 1 & 0 & $2.5$& Large & local & Yes & Varying\\
        S18 & DTHmode24 & Late & 3 & 1 & 0 & $2.5$& Small & local & Yes & shard size\\\hline
        S19 & DTHmode24 & Late & 3 & 1 & 0 & $2.5$& Default & local & No & Turn off\\
        S20 & DTHmode24 & Late & 3 & 1 & 0 & $2.5$& Large & local & No & activated\\
        S21 & DTHmode24 & Late & 3 & 1 & 0 & $2.5$& Small & local & No & REs\\\hline
        M1 & H26 & Late & 3 & 2 & 0 & $20$& Default & local & No & Multiple\\
        M2 & H26 & Late & 3 & 3 & 0 & $20$& Default & local & No & pellets\\
        M3 & H26 & Early & 1 & 2 & 0 & $20$& Default & local & No & \\
        M4 & H26 & Early & 1 & 3 & 0 & $20$& Default & local & No & \\\hline
        M5 & DTHmode24 & Late & 3 & 2 & 0 & $2.5$& Default & local & Yes & Multiple\\
        M6 & DTHmode24 & Late & 3 & 3 & 0 & $2.5$& Default & local & Yes & pellets\\
        M7 & DTHmode24 & Early & 1 & 2 & 0 & $2.5$& Default & local & Yes & \\
        M8 & DTHmode24 & Early & 1 & 3 & 0 & $2.5$& Default & local & Yes & \\\hline
        St1 & H26 & Late & 3 & 1 & 1 & $20$& Default & Shifted & No & Staggered\\
        St2 & H26 & Late & 3 & 2 & 1 & $20$& Default & Shifted & No & injection\\
        St3 & H26 & Late & 3 & 3 & 1 & $20$& Default & Shifted & No & shifted dep.\\\hline
        St1$^*$ & H26 & Late & 3 & 1 & 1 & $20$& Default & Local & No & Staggered\\
        St2$^*$ & H26 & Late & 3 & 2 & 1 & $20$& Default & Local & No & injection\\
        St3$^*$ & H26 & Late & 3 & 3 & 1 & $20$& Default & Local & No & local dep.\\\hline
        St4 & DTHmode24 & Late & 3 & 1 & 1 & $2.5$& Default & Shifted & Yes & Staggered\\
        St5 & DTHmode24 & Late & 3 & 2 & 1 & $2.5$& Default & Shifted & Yes & injection\\
        St6 & DTHmode24 & Late & 3 & 3 & 1 & $2.5$& Default & Shifted & Yes & shifted dep.\\\hline
        St4$^*$ & DTHmode24 & Late & 3 & 1 & 1 & $2.5$& Default & Local & Yes & Staggered\\
        St5$^*$ & DTHmode24 & Late & 3 & 2 & 1 & $2.5$& Default & Local & Yes & injection\\
        St6$^*$ & DTHmode24 & Late & 3 & 3 & 1 & $2.5$& Default & Local & Yes & local dep.\\
    \end{tabular}
    \caption{List of settings for the simulations performed with a $15\,\rm MA$ plasma current with injection before the TQ, which are included in Figs.~\ref{fig:correlations_current}-\ref{fig:correlations_assim}.}
    \label{tab:settings}
\end{table*}

\end{document}